%% file: duohep3.tex
\documentstyle[aps,prd,graphics]{revtex}
\draft 

\def\be{\begin{equation}}
\def\ee{\end{equation}}
\def\bea{\begin{eqnarray}}
\def\eea{\end{eqnarray}}
\def\ba{\begin{array}} %%%%%%%
\def\ea{\end{array}}
\def\bc{\begin{center}}
\def\ec{\end{center}}

\def\physl#1#2#3{Phys. Lett. B#1, #2 (#3)} 
\def\physrev#1#2#3{Phys. Rev. D#1, #2 (#3)}
\def\prl#1#2#3{Phys. Rev. Lett. #1, #2 (#3)}
\def\nucl#1#2#3{Nucl. Phys. B#1, #2 (#3)}

\begin{document}

\include{CelineMacros}

\twocolumn[\hsize\textwidth\columnwidth\hsize\csname
@twocolumnfalse\endcsname 

\title{Light and Heavy Dark Matter Particles} 
\author{C. B\oe hm$^1$, P. Fayet$^2$, J. Silk$^1$} 
\address{$^1$Denys Wilkinson Laboratory, 1 Keble Road, OX1 3RH, 
Oxford, England, UK  
\\
$^2$ Laboratoire de Physique Th\'eorique de l'ENS, UMR 8549, 
24 rue Lhomond, 75231 Paris Cedex 05, France} 
%\thanksref{labo}{\scriptsize UMR 8549, Unit\'e mixte du CNRS et de l'Ecole Normale Sup\'erieure.}
%\date{\today} 
\date{10 November 2003} 
\maketitle

\begin{abstract}
It has recently been  pointed out that the 511 keV emission line detected 
by Integral/SPI  from the bulge of our galaxy 
could be explained by annihilations of light Dark Matter particles into 
$e^+ e^-$. If such a signature is confirmed, then one  might
expect a conflict with the interpretation of very high energy gamma 
rays if they also turn out to be due to Dark Matter annihilations. 
Here, we propose a way to accomodate the existence of both signals
being produced  by 
Dark Matter annihilations through the existence of two stable
(neutral) Dark Matter particles, as is possible in theories inspired 
from $N=2$ supersymmetry.
\end{abstract}


]

\section{Introduction}
Dark Matter particles are expected to produce indirect 
annihilation signatures via production of high  energy $e^+,\bar p, \gamma, \nu$ 
which are potentially detectable \cite{ss}.
An especially intriguing signal  has
recently been observed
with the SPI spectrometer on the INTEGRAL (INTErnational Gamma-Ray
Astrophysics Laboratory) satellite,
which  detected a bright 511 keV
$\gamma$-ray line from the galactic bulge \cite{511}. This detection
turns out to be in good agreement with previous measurements
\cite{previous} and fits reasonably well to a gaussian with full-width
half-maximum of $\sim 9^{\circ}$, with a 2$\sigma$ confidence interval
of $6^{\circ}-18^{\circ}$.

Positrons in the galactic bulge can be emitted by  such 
astrophysical objects as for example hypernovae 
\cite{sn2003}, neutron stars or black holes \cite{compact}
\textit{etc}.  However, whether these sources can be at the origin of
this 511 keV line is still under debate \cite{debate,pohl},
especially in the bulge which consists exclusively of old, low mass stars.
As an alternative explanation, annihilations of light Dark Matter particles
\cite{bens,bf} into $e^+ e^-$ seem to provide a surprisingly  good
explanation for  the Integral measurement, if the dark matter halo
profile is approximated in the centre of the galaxy as $\rho(r)
\propto r^{-\gamma}$ with $\gamma \in [0.4-0.8]$ \cite{betal}.  Such a value
corresponds to a mild cusp  intermediate between the NFW distribution \cite{nfw}
and the profile extrapolated from observations of dwarf and low
surface brightness galaxies \cite{krav}. 
This empirically-derived profile is  consistent with  a  density profile
  that incorporates information on the dark matter contribution
  constrained  by 
gravitational microlensing measurements of the Milky Way
bulge \cite{be}.  Moreover the inferred annihilation cross section is 
concordant with that inferred from the relic density, without any
need to assume a boost factor, due for example to dark matter clumpiness.

One way to determine whether the 511 keV line is due to astrophysical
sources or Dark Matter (DM) annihilations is to seek  a similar
signature from low surface brightness 
dwarf galaxies \cite{smok}.  Those objects are everywhere dark
matter--dominated. Therefore, if the emission line detected in our galaxy is due
to DM annihilations, then one should also detect a 511 keV line from
nearby 
dwarf spheroidals. The flux from the Sagittarius Dwarf Galaxy, in
particular, appears to fall within the expected
sensitivity of Integral/SPI.
% Data in this
%region have already been collected but not analysed as yet. 
It is therefore quite likely that, in the near future, Integral/SPI will 
be able to rule out or confirm the existence of Light Dark Matter (LDM)
particles.

Other experiments seek to detect very high energy gamma rays (\ie of order 
 a hundred GeV) \cite{hegrexp}.  Clearly, if the assumption of LDM 
were to be confirmed, it would seem difficult to accomodate  low as well as 
possible high energy gamma ray signatures (unless, of course, astrophysical
sources appear to be at the origin of the high energy gamma rays).
In fact, the latter may have a natural explanation
within the framework of the (Minimal) Supersymmetric Standard Model.
Annihilations of the lightest neutralino are indeed expected to give
rise to gamma rays of up to several hundred GeV. Whether they would appear a
satisfactory solution or not then depends on the angular and energy
distribution of the observed gamma rays.

However, it is difficult to  accomodate very large neutralino masses (as
required if very energetic gamma rays are detected), with the large
annihilation cross sections needed to comply with the sensitivity of
present/next generation gamma ray experiments, 
unless one invokes the supermassive  black hole
at the centre of our galaxy to steepen the innermost  dark matter cusp
within the zone of influence of the black hole 
and thereby enhance the predicted annihilation rate \cite{bss}. 
The amplitude of this effect is of course dependent on uncertain astrophysical assumptions
about the formation of the supermassive  black hole.

Also, it is worth mentioning that the adoption of
very large neutralino masses requires one to 
invoke the coannihilation mechanism in order to produce a low enough
relic density. However, the conditions for
coannihilation to operate are not always satisfied, as shown in
\cite{bhs}.  In particular, the decay process of the particle with which
the neutralino is supposed to coannihilate (namely the
Next-to-Lightest Supersymmetric Particle, NLSP) may be more efficient 
than the processes that can regenerate this particle.  
As a result, one is left
with the annihilation region (allowing for neutralino masses up to
$\sim$ 100 GeV), plus potentially only a small region of the supersymmetric
parameter space corresponding to the largest neutralino masses (\ie
for which the neutralino is almost mass degenerate with the NLSP and
where the efficiency of the NLSP decay process is significantly reduced).

In this Letter, we point out that it is possible to reconcile
the low and high energy signatures, even if both of them turn
out to be due to \dm annihilations. 
Indeed, existing theories predict  two 
different kinds of Dark Matter particles 
\cite{bf}. 
For example, in a $\,N=1$ supersymmetric framework with mirror fermions 
(somehow reminiscent of $\,N=2$ extended Supersymmetry) \cite{mirror},
one expects the existence of two neutral and stable particles. 
One would be a heavy fermion for example, like the lightest 
neutralino, and the other one a possibly light spin-0 particle. 
Both of them would be neutral and also stable as a result  
of two discrete symmetries (say $R$ and $M$-parities). 

The light particle would be crucial in order to achieve the correct relic
density, while heavy stable particles would be at the origin of very
high energy gamma rays. Light particles would be difficult to detect
in direct Dark Matter detection experiments because of their mass 
but heavy particles would also be hard to find 
due to their small residual number densities.

\section{Relic density, 511 keV line and high energy gamma rays} 

Let us assume that there indeed exists two DM particles. 
The first question to deal with is: how do they share the observed relic density 
\cite{wmap}? 

If the 511 keV emission line is indeed  due to LDM annihilations
in the centre of the Milky Way, then one needs a cross section (times the 
relative velocity, normalised to $c$) of 
about $\sim 10^{-4}$ to $10^{-5}$ pb to fit the
observed flux ($9.9 \times 10^{-4}\,\rm{ph}\,\rm{cm}^{-2}
\,\rm{s}^{-1}$) and the angular distribution of $\gamma$-rays
indicated by Integral/SPI.

On the other hand, it is also necessary to avoid an overproduction of relic LDM
for  the observed low  energy gamma ray flux 
or an overproduction of gamma rays for a given relic
LDM  abundance. Since low DM masses actually
increase the amount of annihilations in our galaxy, and consequently
%To avoid an overproduction of gamma rays, on the other hand, LDM must
%have a $v^2$ suppressed cross section. Since low DM masses actually
%increase the amount of annihilations in our galaxy, and subsequently
the flux of low energy gamma rays that has been well measured by
several experiments, one has to invoke a $v^2$ suppressed annihilation
cross section. One way to achieve this is to assume a scalar particle
coupled to a new light gauge boson \cite{bf}\footnote{ 
This gauge boson is similar to the one introduced 
in \cite{pfu,pfanomalies}. However, in the case we discuss here, it would 
be coupled to Dark Matter and, in fact, would mainly decay into 
two DM particles.}. 

The LDM cross section would then be much larger at early times than
is expected in our galaxy \cite{bens}.  For example, one expects
their annihilation cross section to be about $ 10^{-5}$ pb in the
Milky Way, to explain the 511 keV line, and about 1 pb in the
primordial Universe (since the DM velocity is $\sim 3 \ 10^{-3}$ the
velocity in the primordial Universe).

Given the relationship between the relic density of annihilating DM 
particles and the annihilation cross section in the primordial
Universe, the cosmological parameter associated with these light
particles is then expected to be about
\begin{equation}
\Omega_{\rm{dm}}^{th}h^2 \simeq 0.024 \times
\frac{x_F}{\sqrt{g_{\star}}} \times \left(\frac{\langle\sigma v \rangle_{p}}{1 \U{pb}{}}\right)^{-1}
\end{equation}
where
$x_F$=$m_{\rm{dm}}/T_F\simeq 12-19$ 
for particles in the MeV-GeV range, $g$ and $g_{\star}$
are the numbers of internal and relativistic degrees of freedom,
respectively and $v$ is expressed in units of $c$. Note that
$\langle\sigma v \rangle_{p}$ stands for the primordial annihilation
cross section of LDM. 

For $\sim$ MeV particles, the value of 
$\frac{x_F}{\sqrt{g_{\star}}}$ is $\sim$ 4. Thus one obtains 
$\Omega_{\rm{dm}}^{th}h^2
\simeq 0.095 \ \left(\langle\sigma v \rangle_{p}/(1 \U{pb}{})\right)^{-1}$. 
This has to be compared to $\Omega_{\rm{dm}}h^2 \sim $ 0.1 as found by WMAP.

LDM particles are therefore likely to fit the observed relic density,
leaving little room for another ``weakly annihilating''
species. Of course, this depends on the profile used and the exact
cross section needed to fit the angular resolution of the SPI
detection.
If the annihilation cross section of heavy DM particles
($\langle\sigma v \rangle'$) is larger than $1$ pb at both early
epochs and today\footnote{One would then require constant
annihilation cross sections for heavy particles.}, then these particles
are likely to be subdominant but, on the other hand, they would yield
significant gamma ray fluxes!
This picture would then lead to a situation where one can fit the
observed relic density, explain the nature of the 511 keV line if it
is confirmed and potentially 
predict, at the same time, a non-negligible flux for
high energy gamma rays, even if the dark halo profile is not as cuspy
as the NFW profile.

Assuming that LDM fulfills the relic density condition, one can study
the effect of a large annihilation cross section associated with heavy
particles. Of course, the larger this cross section gets, the less
credible it becomes (as one expects $\sigma \propto 1/m_{\chi}^2$ or
$\propto 1/m_{S}^2$, where $\chi$ and $S$ denote the heavy DM particle
or the exchanged particle  respectively).  However, it is worth
remembering that resonances are possible, in which case 
they would significantly enhance the cross section.

The 511 keV emission line seems to favor a quite flat profile (for
example $\gamma \sim 0.4$). If so, then the annihilation cross section
needed may appear extremely large. In fact, this would
be probably too large compared
to what could be realistically expected in a particle physics model
unless, perhaps, new kinds of interactions are invoked.
On the other hand, even a quite mild dark halo profile (\ie $\gamma
\sim 0.8$) would yield a significant amount of high energy gamma rays
if the annihilation cross section associated with heavy DM particles
turns out to be large but still ``realistic'' (\ie enhanced by 
resonance effects for example). Such a scenario would then be in
agreement with the \dm cosmological parameters measured by WMAP as well
as with low and high energy annihilation signatures (if they were both
confirmed).

\section{A theory with two stable Dark Matter particles} 

A very important requirement regarding any \dm candidate is its stability.  
In a Standard Supersymmetric framework, the neutralino 
(absolute) stability is obtained through a $Z_2$ discrete 
symmetry, called $R$-parity \cite{pf1}.  This symmetry 
implies the decay of a supersymmetric particle into another one, plus 
standard model particles. This is what prevents 
the lightest supersymmetric particle (LSP) from decaying.

Since we are now looking for two stable particles, it is particularly 
interesting to start working within the framework 
of $N=2$ extended supersymmetric theories, as they may naturally allow 
for two $Z_2$ discrete symmetries, and may ultimately lead 
to two new distinct DM candidates. 
We shall denote these two discrete symmetries as $R$ and $M$ parities. 

Note that the particles which matter for our purpose 
(say mirror leptons/quarks and new spin-0 states that couple these 
mirror particles to ordinary ones \cite{mirror}) may be kept even after one  
abandons the full \hbox{$N=2$} supersymmetry in favor of a $N=1$ theory.

\subsection{Two discrete symmetries}

As mentioned before, $N=1$ supersymmetric theory may admit a 
discrete $\,Z_2\,$ symmetry. The latter can be seen, in fact, 
as a remnant of a continuous $\,R$-symmetry ($\,R_p=(-1)^R$).  

Thus, quite similarly to the case $N=1$, an extended $\,N=2\,$ theory, 
for which there are now two  supersymmetry generators 
\be
Q = \left(\! \ba{c} Q^1_{\,L} \vspace{1mm}\\ Q^2_{\,L} 
\ea \!\right), 
\ee
may admit a $\,SU(2)\times U(1)\,$ global $R$-symmetry, 
that acts on the doublet $Q$. 

Such a global $R$-symmetry allows one to act 
independently on the two supersymmetry generators
$Q_1$ and $Q_2$. One can therefore define two distinct 
$R$-parity symmetry operators, $R_{1\,p}$ and $R_{2\,p}$; 
the usual $\,R$-parity symmetry  corresponding to the product 
\begin{equation}
R_p\ =\ R_{1\,p} \ R_{2\,p}\ \ .
\end{equation}

By abandoning the full $N=2$ supersymmetry, one 
singles out one of the two supersymmetry 
generators (say $Q_1$) and redefines one of the two discrete symmetries 
(say $R_{2\,p}$) as the $M$-parity symmetry. 

We are then left with two discrete symmetries: the  
$M$-parity that can be seen as a $Z_2$ remnant of a 
continuous $U(1)$ symmetry ($M$) acting globally on $N=1$ 
superfields, and the ``standard'' $R$-parity. 
Both of them are kept intact after 
spontaneous breaking of the electroweak symmetry.

\subsection{Particle spectrum from extended supersymmetry}

We now have to specify what is the expected spectrum from 
extended supersymmetry and, more precisely, which particle 
could be the LMP. 

When $N\!=\!2$ extended supersymmetric theories are 
formulated in terms of  $N\!=\!1$ superfields,  
ordinary gauge superfields are accompanied by additional ($N\!=\!1$) 
``chiral gauge superfields''.  Together, they describe 
\hbox{$N\!=\!2$} massless gauge multiplets containing, for example, 
two gluino octets, two photinos, \textit{etc.}, 
as well as two color-octets of spin-0 gluons, 
two spin-0 photons, \textit{etc.} \cite{mirror}.
(These new ``scalar gauge  fields'' may also appear as originating 
from the fifth and sixth components  of higher-dimensional gauge fields 
$\,V^{\hat\mu}$, \,in a six-dimensional spacetime\,\cite{6d}.)

Matter (and also Higgs) chiral superfields systematically 
occur in pairs so as to describe $\,N=2\,$ ``hypermultiplets''. 
Since quark and lepton fields belong to left-handed electroweak
doublets and right-handed singlets, one has  to introduce 
{\it \,mirror\,} particles  which belong to right-handed 
{\it \,doublets\,} and left-handed {\it \,singlets\,}.
This also necessitates however the introduction of appropriate 
symmetry breaking mechanisms, allowing one to reduce 
the full $N=2$ supersymmetry down to $N=1$.

Mirror quarks and leptons ($q_M$ and $l_M$)
may then acquire large masses through large Yukawa couplings 
with electroweak Higgs doublets 
(namely $H_1$ and $H_2$ in the standard supersymmetry framework).
They would have evaded past accelerator searches if heavier 
than a few hundred GeV's, but may still show up at LHC.

The breaking of the ($N=2$) supersymmetry may be elegantly obtained by 
demanding periodic and antiperiodic boundary conditions for ordinary 
$R$-even particles and their $R$-odd 
superpartners, respectively -- in which case the masses of the (lowest-lying)
gravitinos, gluinos and photinos,
which fix the energy scale at which supersymmetric particles should start 
to show up, would be given, in the simplest case 
and up to radiative correction effects, by 
$
m_{3/2}=m_{1/2}=\frac{\pi\,\hbar}{L\,c}=\frac{\hbar}{2\,R\ c}\ .
$
$L$ is the size of the extra dimension responsible for supersymmetry 
breaking.
%%%(or of the corresponding ``radius'' $R$).
This led us to consider the possibility of relatively ``large'' extra 
dimensions, associated with a compactification scale that could then be 
as ``low'' as $\,\sim $ TeV scale\,\cite{compac}.

It is quite conceivable that the new spin-0 states,
as well as mirror lepton and quark fields, may only manifest themselves  
at the compactification scale. However, whether the extra spin-0 components 
of gauge fields actually show up 
or not in the low-energy theory depends on the details of the mechanism 
that should be responsible for the breaking of the extended supersymmetry
(and/or the compactification of the extra space dimensions). 
Since this mechanism is unknown, one can still discuss the possibility that 
they appear in the low-energy theory, much below the compactification scale.

The new spin-0 bosons -- originally appearing as extra degrees of freedom 
for spin-1 gauge fields  -- will then have Yukawa couplings 
relating quarks and leptons to their mirror partners. They 
should lead to new decay modes ($M$-parity invariant) like
\be
q_M\ (\hbox{or} \ l_M) \ \rightarrow\ q\ (\hbox{or}\ l)\ + \
\hbox{\small
neutral spin-0 particle}\ \ .
\ee

Note that $M$-parity is equal to $+1$ for ordinary and 
Supersymmetric Standard Model particles 
(including those described by the doublet Higgs superfields 
$H_1$ and $H_2$). It is equal to $-1$, on the other hand, 
for mirrors, their superpartners, the new 
spin-0 particles, the new inos (second octet of gluinos, 
additional charginos and neutralinos) and finally also for the extra 
Higgs bosons 
which appear as the trace of the underlying extended supersymmetry.

The lightest of the new $M$-odd particles, that remains 
at the end of such decay chains, is then expected to be stable. 

\subsection{A Light Dark Matter candidate} 

In the scenario described above, one ends up with a new neutral and 
stable ($M$-odd) particle that can play the r\^ole of LDM candidate. 

This particle could be a scalar (that couples 
ordinary to mirror particles). In which case, it 
is expected to annihilate in pairs through mirror fermion exchanges, 
as discussed in \cite{bf}. However, such a scenario is disfavored 
by the 511 keV line observation (which requires $v^2$ suppressed 
annihilation cross section), unless the associated cross section 
turns out to be extremely small in the primordial universe 
(namely $\lesssim 10^{-31} \U{cm}{3} \U{s}{-1}$, which requires 
heavy mirror fermions and probably chiral couplings).  

The LDM particle may also annihilate through possible couplings to a 
spin-1 $ \ U$ boson,
if the extra $U(1)$ symmetry generator includes a contribution 
involving the continuous $U(1)$ $M$-symmetry generator  
\footnote{Such a gauging of an extra $U(1)$ not commuting 
with the $N=2$ supersymmetry may however
generate anomalies (at least superficially).}. 
In this case, one would obtain an attractive LDM candidate 
that could explain the low energy signature without being in conflict 
with a signal at higher energies.

\section{Summary}

Supersymmetric theories originating from $N=2$ have the pleasant 
feature of providing 
two discrete symmetries, that we denote $R$ and $M$-parities.  
When they are conserved, one expects that both the 
LSP (Lightest Particle invariant under $R$-parity) and LMP 
(Lightest Particle of the spectrum invariant under $M$-parity) become 
absolutely stable. Such a scenario would then offer two DM candidates: 
one would be the $R$-odd LSP and the other one, the $M$-odd LMP.  

Among the new $M$-odd particles, one finds spin-0 states.  
The case where one of them turns out to be the lightest of the 
$M$-odd spectrum appears particularly appealing as it could play 
the r\^ole of Light Dark Matter candidate (with a mass of a few MeV-100 MeV). 
Provided one introduces a new light gauge boson $U$, 
one could  evade low energy gamma ray constraints \cite{bf} and also 
explain the 511 keV emission line (if it is seen to be due to DM annihilations)
\footnote{There is no need to introduce a gauge boson for $\sim$ 
100 MeV but this case appears borderline regarding 
the 511 keV line, the relic density criterion and the nucleosynthesis constraint.}.

To avoid an unwanted production in $Z$ decays, this spin-0 particle 
should have no or negligible direct couplings to the $Z$. 
This would restrict potential LMP candidates 
to e.g. a spin-0 photon, 
or companion of the weak hypercharge gauge field $B^\mu$. 

In such a scenario, one would for example end up with the 
``standard'' (lightest) neutralino as the $R$-odd LSP  and 
a new neutral spin-0 particle,  as $M$-odd LMP.

The new $M$-odd spin-0 states are expected to couple quarks and
leptons to {\it \,mirror\,} partners.  
However, if the latter are very heavy (a few hundred GeV for example),
one does not expect a significant contribution to low energy gamma rays. 
(The associated annihilation cross section would indeed be dominated by a 
constant term but this one is not expected to yield any significant 
contribution as it would be lower than $10^{-5}$ pb. )
On the other hand, one expects high energy gamma rays from 
the annihilations of heavy DM particles for which $\sigma v' > 1$ pb. 

This scenario finally evades several astrophysical and particle physics 
constraints including  too large a contribution to the muon and 
electron $g-2$. On the other hand, a gauge boson is expected to bring an 
extra contribution to the $\nu-e$ 
elastic scattering cross section at low energy. 
The latter has been measured by two experiments (LAMPF and LSND).  
No significant deviations have been found \cite{lampf,lsnd}. Therefore, 
it is necessary to either impose a dissymmetry between 
the coupling of the new gauge boson to DM and its coupling to standard model 
particles or, alternatively,  to suppress its coupling to neutrinos
(e.g. by having the $q$ and $\,l$ contribution 
to the $U$ current proportional to the electromagnetic current,
as discussed in \cite{pfanomalies}).

Of course, there may exist other ways out to achieve a $v^2$ 
suppressed cross section for LDM. We do not discuss them in this letter. 
 
By having two kinds of DM particles, we then fulfill the correct 
relic density, potentially 
explain the 511 keV emission line detected by Integral/SPI and 
simultaneously predict high energy gamma rays. 

This would turn out to be particularly interesting either 
if the smoking gun proposed in \cite{smok} reveals to be in favor of LDM 
and if high energy diffuse gamma rays were identified as coming from 
\dm annihilations, or simply if the high energy gamma rays 
turned out to be associated with a \dm annihilation cross section that is 
much larger compared to what is needed to achieve the correct relic density.

\end{document}

%% file: CelineMacros.tex
\newcommand{\U}[2]{\,{\rm #1}^{#2}}

\def\adec {a_{dec(dm-\nu)}}
\def\gdec {\tilde{\Gamma}_{dec(dm-\nu)}}
\def\Tdec {T_{dec(dm-\nu)}}
\def\tdec {t_{dec(dm-\nu)}}
\def\dm {Dark Matter }
\def\dmsb {Dark Matter}
\def\cs {\sigma_{(dm-\nu)}} 
\def\bel {b_{el(dm-\nu)}}
\def\nt {\tilde{n}_{\nu}}
\def\mdm {m_{dm}}
\def\cm3 {\mbox{cm}^3 \, \mbox{s}^{-1}}

\def\beqn{\begin{eqnarray}}
\def\eqn{\end{eqnarray}}
\def\ie {\textit{i.e.} }
\def\eg {\textit{eg.} }
\def\etal {\textit{et al.} }
\def\vs {\vspace{0.2cm}}
\def\ghost#1 { }

\def\stau { \tilde{\tau}}
\def\mstau { m_{\tilde{\tau}}}
\def\nstau { n_{\tilde{\tau}}}
\def\nstaue {n_{\tilde{\tau}, \, e}}
\def\chio { \chi_1^0}
\def\mchi { m_{\chi}}
\def\nchi { n_{\chi}}
\def\nchie { n_{\chi, \, e}}
\def\ntau { n_{\tau}}
\def\ntaue {n_{\tau, \, e}}
\def\tbd {three body decay rate }
\def\dbd {two body decay rate }
\def\fbd {four body decay rate }

\def\ubar#1{\overline{u}_{#1}}
\def\u#1{{u}_{#1}}
\def\v#1{{v}_{#1}}
\def\vbar#1{\overline{v}_{#1}}
\def\hs{\hspace{0.15cm}}
\def\qp{p}
\def\dslash#1#2{ \mbox{$#1$ \kern-0.9em \slash \kern0.2em}_{#2} }

%% file: duohep3.bbl
\begin{thebibliography}{99} 

\bibitem{ss}
J.~Silk and M.~Srednicki,
\prl{53}{624}{1984}.

\bibitem{511}
J.~Knodlseder, {\it et.al.},
Accepted for publication in A\&A,
arXiv:astro-ph/0309442;
P.~Jean {\it et al.},
%``Early SPI/INTEGRAL measurements of galactic 511 keV line emission
arXiv:astro-ph/0309484.
%%CITATION = ASTRO-PH 0309484;%%

\bibitem{previous}
D.~D.~Dixon {\it et al.},
%%``Skymapping with OSSE via the Mean Field Annealing Pixon Technique,''
arXiv:astro-ph/9703042;
%%%CITATION = ASTRO-PH 9703042;%%
P.~A.~Milne, J.~D.~Kurfess, R.~L.~Kinzer and M.~D.~Leising,
%%``Comparative Studies of Line and Continuum Positron Annihilation 
arXiv:astro-ph/0106157.
%%%CITATION = ASTRO-PH 0106157;%%

\bibitem{sn2003}
S.~E.~Woosley and A.~Heger,
%``The Light Curve of the Unusual Supernova SN 2003dh,''
arXiv:astro-ph/0309165.
%%CITATION = ASTRO-PH 0309165;%%

\bibitem{compact}
R.~E.~Lingenfelter and R.~Ramaty,
Positron-Electron Pairs in Astrophysics, eds. M.~L.~Burns, A.~K.~Harding and R.~Ramaty, AIP Conference Proceedings, 267.


\bibitem{debate}
P.~A.~Milne, L.~S.~The and M.~D.~Leising,
%``Late Light Curves of Type Ia Supernovae,''
arXiv:astro-ph/0104185.
%%CITATION = ASTRO-PH 0104185;%%
P.~A.~Milne, J.~D.~Kurfess, R.~L.~Kinzer and M.~D.~Leising,
%``Supernovae and Positron Annihilation,''
New Astron.\ Rev.\  46, 553 (2002)
%[arXiv:astro-ph/0110442], 
and references therein.
%%CITATION = ASTRO-PH 0110442;%%


\bibitem{pohl}
M.~Pohl,
%``An annihilation fountain at the Galactic center?,''
arXiv:astro-ph/9807268.
%%CITATION = ASTRO-PH 9807268;%%



\bibitem{bens}
C.~Boehm, T.~A.~Ensslin and J.~Silk,
%``Are light annihilating dark matter particles possible?,''
arXiv:astro-ph/0208458.
%%CITATION = ASTRO-PH 0208458;%%


\bibitem{bf}
C.~Boehm and P.~Fayet,
%``Scalar dark matter candidates,''
arXiv:hep-ph/0305261.
%%CITATION = HEP-PH 0305261;%%

\bibitem{betal}
C.~Boehm {\it et al.}, 
%``MeV Dark Matter: Has It Been Detected?''
arXiv:astro-ph/0309686.
%%CITATION = ASTRO-PH 0309686;%%


\bibitem{nfw}
%J.~F.~Navarro, C.~S.~Frenk and S.~D.~White,
%%``The Structure of Cold Dark Matter Halos,''
%Astrophys.\ J.\  {\bf 462}, 563 (1996)
%[arXiv:astro-ph/9508025];
%%CITATION = ASTRO-PH 9508025;%%
J.~F.~Navarro, C.~S.~Frenk and S.~D.~White,
%``A Universal density profile from hierarchical clustering,''
Astrophys.\ J.\  490, 493 (1997).
%%CITATION = ASJOA,490,493;%%

\bibitem{krav}
A.~V.~Kravtsov, A.~A.~Klypin, J.~S.~Bullock and J.~R.~Primack,
%``The Cores of Dark Matter Dominated Galaxies: theory vs. observations,''
arXiv:astro-ph/9708176.
%%CITATION = ASTRO-PH 9708176;%%


\bibitem{be}
J.J. Binney,  N.W. Evans,
%``Cuspy Dark-Matter Haloes and the Galaxy''
astro-ph/0108505.
%%CITATION = ASTRO-PH 0108505;%%

\bibitem{smok}
D. Hooper \textit{et al},  
%``MeV Dark Matter In Dwarf Spheroidals: A Smoking Gun?''
astro-ph/0311150.  
%%CITATION = ASTRO-PH 0311150;%%

\bibitem{hegrexp} 
A. Djannati-Atai, for the H.E.S.S. collaboration 
``First results from southern hemisphere extragalactic 
observations with H.E.S.S.'', Presentation at ICRC 2003, Tsukuba;
R. Enomoto (ICRR), S. Hara (TIT), CANGAROO collaboration, 
%``Design Study of CANGAROO-III, Stereoscopic Imaging Atmospheric Cherenkov Telescopes for sub-TeV Gamma-ray''
Astropart.Phys. 16 (2002) 235-244.
%%CITATION = ASTRO-PH 0107578;%%

\bibitem{bss}
G. Bertone, G. Sigl and J. Silk, 
Mon.Not.Roy.Astron.Soc. 337 (2002) 98.
%%CITATION =   ASTRO-PH 0203488; %%

\bibitem{bhs}
C. Boehm, D. Hooper and J. Silk, in preparation.  
% ``Killing the coannihilation mechanism?''

\bibitem{mirror} P. Fayet, \nucl{149}{137}{1979}; Proc. 1980 Karpacz 
School (Harwood, 1981) p.115; \physl{142}{263}{1984}; 
\nucl{246}{89}{1984}.


\bibitem{wmap}
C.~L.~Bennett {\it et al.},
``First Year Wilkinson Microwave Anisotropy Probe (WMAP)'',  
arXiv:astro-ph/0302207.
%%CITATION = ASTRO-PH 0302207;%%


\bibitem{pfu} P. Fayet, \physl{95}{285}{1980}; \nucl{187}{184}{1981}. 

\bibitem{pfanomalies} P. Fayet, \nucl{347}{743}{1990}.

\bibitem{pf1} P. Fayet, \physl{69}{489}{1977}; 
G. Farrar and P. Fayet, \physl{76}{575}{1978}.

\bibitem{6d} P. Fayet, \physl{159}{121}{1985}.

\bibitem{compac} P. Fayet, \nucl{263}{649}{1986};
Proc. 2nd Nobel Symp. on El. Part. Physics (Marstrand, Sweden, 1986), 
Physica Scripta T 15 (1987) 46.


\bibitem{lampf} R. C. Allen \textit{et al.}, \physrev{47}{11}{1993}.

\bibitem{lsnd} LSND coll., \physrev{63}{112001}{2001}.


%%%\bibitem{osse}
%G.~Hasinger, A\&A Suppl. 120, 607 (1996), and references therein;
%P.~Sreekumar, F.~W.~Stecker and S.~C.~Kappadath,
%%``The Extragalactic Diffuse Gamma-Ray Emission,''
%arXiv:astro-ph/9709258.
%%CITATION = ASTRO-PH 9709258;%%
%D.~D.~Dixon {\it et al.},
%%``Skymapping with OSSE via the Mean Field Annealing Pixon Technique,''
%arXiv:astro-ph/9703042;
%%%CITATION = ASTRO-PH 9703042;%%
%P.~A.~Milne, .~D.~Kurfesss, .~L.~Kinzers, .~D.~Leisings, J.~D.~Kurfess, R.~L.~%Kinzer and M.~D.~Leising,
%%``Comparative Studies of Line and Continuum Positron Annihilation 
%arXiv:astro-ph/0106157.
%%%CITATION = ASTRO-PH 0106157;%%


%%%\bibitem{leew}
%B.~W.~Lee and S.~Weinberg,
%``Cosmological Lower Bound On Heavy-Neutrino Masses,''
%Phys.\ Rev.\ Lett.\  {\bf 39}, 165 (1977);
%%CITATION = PRLTA,39,165;%%

%%%\bibitem{detection}
%W.~N.~Johnson and R.~C~.Haymes,
%Astrophys.\ J.\  {\bf 184}, 103 (1973).

%%%\bibitem{id}
%M.~Leventhal, C.~J.~MacCallum and P.~D.~Stang,
%Astrophys.\ J.\  {\bf 225}, L11 (1978).


%%%\bibitem{stars}
%R.~Ramaty, B.~Kozlovsky and R.~E.~Lingenfelter,
%Astrophys.\ J.\ Suppl.\ {\bf 40}, 487 (1979).

%%%\bibitem{ism}
%B.~Kozlovsky, R.~E.~Lingenfelter and R.~Ramaty,
%Astrophys.\ J.\  {\bf 316}, 801 (1987).

%%%\bibitem{pulsars}
%P.~A.~Sturrock,
%Astrophys.\ J.\  {\bf 164}, 529 (1971).

%\bibitem{rate}
%F.~Matteucci, D.~Romano and P.~Molaro,
%``Light and Heavy Elements in the Galactic Bulge,''
%arXiv:astro-ph/9810125.
%%CITATION = ASTRO-PH 9810125;%%



%%%\bibitem{casse}
%M.~Casse, {\it et al.},in preparation.

%%%\bibitem{starbursts}
%J.~Bland-Hawthorn and M.~Cohen,
%``The Large-scale Bipolar Wind in the Galactic Center,''
%Astrophys.\ J.\  {\bf 582}, 246 (2003)
%[arXiv:astro-ph/0208553].
%%CITATION = ASTRO-PH 0208553;%%

%%%\bibitem{fountain}
%C.~D.~Dermer and J.~G.~Skibo,
%``Annihilation Fountain in the Galactic Center Region,''
%arXiv:astro-ph/9705070.
%%CITATION = ASTRO-PH 9705070;%%

%C.~Pryke, {\it et al.},
%N.~W.~Halverson, E.~M.~Leitch, J.~Kovac, J.~E.~Carlstrom, W.~L.~Holzapfel and %M.~Dragovan,
%``Cosmological Parameter Extraction from the First Season of 
%Astrophys.\ J.\  {\bf 568}, 46 (2002);
%%CITATION = ASTRO-PH 0104490;%%
%S.~Perlmutter {\it et al.}  [Supernova Cosmology Project Collaboration],
%``Measurements of Omega and Lambda from 42 High-Redshift Supernovae,''
%Astrophys.\ J.\  {\bf 517}, 565 (1999);
%%CITATION = ASTRO-PH 9812133;%%
%S.~Burles, K.~M.~Nollett, J.~W.~Truran and M.~S.~Turner,
%``Sharpening the predictions of big-bang nucleosynthesis,''
%Phys.\ Rev.\ Lett.\  {\bf 82}, 4176 (1999);
%%CITATION = ASTRO-PH 9901157;%%
%M.~Fukugita, C.~J.~Hogan and P.~J.~Peebles, 
%``The Cosmic Baryon Budget,''
%Astrophys.~J. {\bf 503}, 518 (1998);
%%CITATION = ASTRO-PH 9712020;%
%M.~Davis, G.~Efstathiou, C.~S.~Frenk and S.~D.~White,
%``The Evolution Of Large-Scale Structure In A Universe Dominated
%Astrophys.\ J.\  {\bf 292}, 371 (1985).
%%CITATION = ASJOA,292,371;%%

%%%\bibitem{neutralino}
% J.~R.~Ellis, J.~S.~Hagelin, D.~V.~Nanopoulos, K.~A.~Olive and M.~Srednicki,
  %``Supersymmetric Relics From The Big Bang,''
%  Nucl.\ Phys.\ B {\bf 238}, 453 (1984);
  %%CITATION = NUPHA,B238,453;%%
% H.~Goldberg,
  %``Constraint On The Photino Mass From Cosmology,''
%  Phys.\ Rev.\ Lett.\  {\bf 50}, 1419 (1983).
  %%CITATION = PRLTA,50,1419;%%


%%%\bibitem{pfu} P. Fayet, \nucl{187}{184}{1981}; \physl{171}{261}{1986}.
%%%\bibitem{bsp} P. Fayet,  \physl{227}{127}{1989}; \nucl{347}{743}{1990}.
%%%\bibitem{griest} K. Griest, \physrev{38}{2357}{1988}, 
%Erratum-ibid. D39, 3802 (1989);  
%M. Drees, M. M. Nojiri, \physrev{47}{376}{1993}. 
%\bibitem{egret} G. Hasinger, A\&A Supp.  120,
%607 (1996), and references therein; P. Sreekumar \textit{et al.}, \apj{494}{523}{1998}.

% B.~C.~Allanach, A.~Dedes and H.~K.~Dreiner,
%  %``Bounds on R-parity violating couplings at the weak scale and at the 
%  Phys.\ Rev.\ D {\bf 60}, 075014 (1999).
%   %%[arXiv:hep-ph/9906209].
%  %%CITATION = HEP-PH 9906209;%%


%%%\bibitem{longair}
%M.~S.~Longair,
%High-Energy Astrophysics, Vol.~2, Ch.~19 (1981). 

%%%\bibitem{line}
%C.~D.~Dermer and R.~J.~Murphy,
%``Annihilation Radiation in the Galaxy,''
%arXiv:astro-ph/0107216.
%%CITATION = ASTRO-PH 0107216;%%


%%%\bibitem{superfield} B. Delamotte, F. Delduc, P. Fayet, \physl{176}{409}{1986}; 
%B. Delamotte, P. Fayet, \physl{195}{563}{1987}.

%%%\bibitem{compac} P. Fayet, \nucl{263}{649}{1986};
%Proc. 2nd Nobel Symp. on El. Part. Physics (Marstrand, Sweden, 1986), 
%Physica Scripta T 15 (1987) 46.

%\bibitem{smm}
%K.~Watanabe, {\it et al.},
%Proceedings of the 4th Compton Symposium, Williamsburg, Virginia (1997).

%%%\bibitem{osse}
%D.~D.~Dixon {\it et al.},
%%``Skymapping with OSSE via the Mean Field Annealing Pixon Technique,''
%arXiv:astro-ph/9703042;
%%%CITATION = ASTRO-PH 9703042;%%
%P.~A.~Milne, .~D.~Kurfesss, .~L.~Kinzers, .~D.~Leisings, J.~D.~Kurfess, R.~L.~%Kinzer and M.~D.~Leising,
%%``Comparative Studies of Line and Continuum Positron Annihilation 
%arXiv:astro-ph/0106157.
%%%CITATION = ASTRO-PH 0106157;%%

%\bibitem{asca}
%K.~C.~Gendreau, 
%Ph.~D dissertation, Massachussets Institute of Technology, Cambridge, Massachu%setts (1995).

%%%\bibitem{nucleosynthesis}
%P.~McDonald, R.~J.~Scherrer and T.~P.~Walker,
%``Cosmic microwave background constraint on residual annihilations of  relic particles,''
%Phys.\ Rev.\ D {\bf 63}, 023001 (2001)
%[arXiv:astro-ph/0008134];
%%CITATION = ASTRO-PH 0008134;%%
%J.~A.~Frieman, E.~W.~Kolb and M.~S.~Turner,
%``Eternal Annihilations: New Constraints On Longlived Particles From Big Bang Nucleosynthesis,''
%Phys.\ Rev.\ D {\bf 41}, 3080 (1990).
%%CITATION = PHRVA,D41,3080;%%

%%%\bibitem{hearty}
%C. Hearty, {\it et al.},
%Phys.\ Rev.\ D {\bf 39}, 3207 (1989).

\end{thebibliography}
